\documentclass[preprint]{aastex}

\def\spose#1{\hbox to 0pt{#1\hss}}
\def\ltsim{\mathrel{\spose{\lower 3pt\hbox{$\mathchar"218$}}
     \raise 2.0pt\hbox{$\mathchar"13C$}}}
\def\gtsim{\mathrel{\spose{\lower 3pt\hbox{$\mathchar"218$}}
     \raise 2.0pt\hbox{$\mathchar"13E$}}}

\shorttitle{Stepinski, Malhotra \& Black}
\shortauthors{The Upsilon Andromedae System}

\begin{document}

\title{The Upsilon Andromedae System: Models and Stability}


\author{Tomasz F. Stepinski \quad({\tt tom@lpi.usra.edu})}
\author{Renu Malhotra \quad({\tt renu@lpi.usra.edu}) }
\author{and}
\author{David C. Black \quad({\tt black@lpi.usra.edu}) }

\author{}

\affil{Lunar and Planetary Institute, 3600 Bay Area Blvd., Houston, TX
77058}

\begin{abstract}
Radial velocity observations of the F8 V star $\upsilon$ Andromedae taken 
at Lick and at Whipple Observatories have revealed evidence of three
periodicities in the line-of-sight velocity of the star.
These periodicities have been interpreted as evidence for at least three
low mass companions (LMCs) revolving around $\upsilon$ Andromedae.
The mass and orbital parameters inferred for these companions raise questions
about the dynamical stability of the system.
We report here results from our independent analysis of the published radial 
velocity data as well as new unpublished data taken at Lick Observatory.
Our results confirm the finding of three periods in the data.
Our best fits to the data, on the assumption that these periods arise from the
gravitational perturbations of companions in keplerian orbits, is also
generally in agreement, but with some differences, from the earlier findings.
We find that the available data do not constrain well the orbital eccentricity
of the middle companion in a three-companion model of the data.  
We also find that in order for our best-fit model to the Lick data to be
dynamically stable over the lifetime of the star ($\sim\!2$ billion years),
the system must have a mean inclination to the plane of the sky greater than
13 degrees.  The corresponding minimum inclination for the best fit to the
Whipple data set is 19 degrees.  These values imply that the maximum mass
for the outer companion can be no greater than about 20 Jupiter masses.
Our analysis of the stability of the putative systems also places constraints
on the relative inclinations of the orbital planes of the companions.
We comment on global versus local (i.e., method of steepest descent) 
means of finding best-fit orbits from radial velocity data sets. 
\end{abstract}


\section{Introduction}
Radial velocity observations of several hundred nearby main-sequence stars 
have resulted in the detection to date of roughly forty companions with minimum 
or projected masses (i.e., $m \sin i$) less than 80 $M_{\rm J}$, 
where $M_J$ is the mass of Jupiter \citep{mar98, mar99, vog00, may97}.
Nearly thirty of these companions have $m \sin i \ltsim 10$ $M_{\rm J}$.
The evidence to date typically suggests only one companion per star; however,
the data from some of the observational studies have suggested that other
companions may be present in some cases but will require longer time bases
for the observational record before firm conclusions can be established
\citep{cum99}.

Evidence for the presence of multiple low mass companions (LMCs) to a star
would be significant for several reasons. A key reason is that it would suggest,
at least superficially, a similarity of such a system to our planetary system,
and by extension, planetary systems in general. 

The first strong evidence for multiple LMCs to a single star,
Upsilon Andromedae ($\upsilon$ And), was reported recently by
\citet{but99} (hereafter referred to as B99).
The data presented in that paper are from two independent studies,
one conducted at the Lick Observatory and the other conducted at the
Whipple Observatory using the Advanced Fiber-Optic Echelle (AFOE) Spectrograph. 

Earlier observations of $\upsilon$ And \citep{but97} had detected a 
periodicity in the radial velocity data that indicated the presence of a
companion with an orbital period of 4.6 days and a projected mass of
$\sim\!0.7\ M_{\rm J}$. Those authors noted that the data also contained
``evidence for variability in the gamma velocity with timescale of about 2 yr."
The newer observations (B99) reveal additional periodicities,
one in excess of 1200 days (1269 days for the Lick data and 1481 days
for the AFOE data), as well as one with $\sim\!240$ days. 

B99 have modeled these periodicities as arising from the presence of
three LMCs in keplerian orbits about $\upsilon$ And.
The eccentricities of the orbits determined by B99 are, in order of increasing
orbital period, 0.042, 0.23, and 0.36. This three-companion model for the 
observations raises interesting challenges in understanding the formation
and evolution of that system and its possible relationship to systems such
as the Solar System. Particular challenges relate to the dynamical stability
of such a system and what constraints it might place on physically
realizable companion systems, and to how a system consisting of at least
three relatively massive objects could form around a star with the orbital
structure that is suggested by the data.

In an effort to explore these challenges in more detail,
we re-examine here the published radial velocity data on $\upsilon$ And, 
as well as data taken subsequent to announcement of the results and kindly 
provided to us by G.~Marcy.  Section 2 summarizes the radial velocity data that 
are used in our analyses.  Models for analyzing the radial velocity data are 
discussed in Section 3.  The methods that we used for fitting the data and the 
procedures for assessing the merit of those fits are described in Section 4.
The best-fit models, assuming that the periodicities are due to companions
(i.e., that the model for describing the data is one comprised of a
superposition of Keplerian motion) are presented in Section 5.
We examine the dynamical stability of candidate model systems in Section 6.
A summary of our results and conclusions regarding the possible nature of
the $\upsilon$ And system are given in Section 7.

\section{Radial velocity data}
We use three different data sets to perform our analysis.
The first data set was collected at Lick.
It contains 89 observations of $\upsilon$ And made between September 1987
and March 1999 as part of the Lick survey.
We refer to this data set as the original Lick data.
The second data set was collected by the AFOE planet search program and
contains 52 observations of $\upsilon$ And made between September 1994
and February 1999. We refer to this data set as the AFOE data.
Original Lick data and AFOE data are published by B99.
Details about these data sets can be found therein.
The third data set, referred to as the new Lick data, has
been provided to us by G.~Marcy. This contains 118 observations of  
$\upsilon$ And collected at Lick and comprises 29 observations 
made between June and August of 1999 in addition to the original
89 observations. Note that in this data set the radial velocities
for the original 89 observations have been revised to reflect
improvement in the data reduction technique.

All three data sets have the same form:
each record is a triplet $(t, V, \sigma)$,
where $t$ is the time of observation (in Julian days),
$V$ is an unaccounted for component of the star's radial velocity
(hereafter referred to as radial velocity) in (m\,s$^{-1}$),
and $\sigma$ is a measurement error in (m\,s$^{-1}$).
Bulk properties of radial velocities are consistent among the three
data sets. The range of $V$ is $-177$\,m\,s$^{-1}$ to $165$\,m\,s$^{-1}$,
the mean value of $V$ is $-11$\,m\,s$^{-1}$ to $-5$\,m\,s$^{-1}$,
and the standard deviation is $71$\,m\,s$^{-1}$ to $82$\,m\,s$^{-1}$.
Measurement errors are of the order of 10\,m\,s$^{-1}$,
generally smaller for the Lick data than for the AFOE data.

It is useful to think about the radial velocity data from
$\upsilon$ And as a time series, and to pretend that we have no
a priori insight into the mechanism that produces it. The first step
is to calculate the signal's frequency spectrum. Fig.~1 (left column) shows
the frequency spectrum for all three data sets.
Because the observations are not evenly spaced, the frequency spectrum
cannot be obtained by means of the FFT, instead we used the Lomb-Scargle
periodogram technique \citep{lom76, sca82, bla82}
to obtain standard, zero mean periodograms.

These spectra indicate the existence of periodic components in the radial
velocity signal from $\upsilon$ And. Spectral features common to all
three data sets exist. The most prominent are peaks at $\sim\!4.617$\,d,
$\sim\!500$\,d, and $\sim\!1200$--1500\,d.
However, not all significant peaks present in the frequency spectrum
actually correspond to real periodicities.
A simple test for the reality of periodicities indicated by the frequency
spectrum is to fold the signal with suspected periods. Only folds with actual
periodicities yield coherent patterns. The fold test provides definitive
affirmation, but, in the presence of multiple periodicities it does not
necessarily provide definitive disaffirmation. 
Using the fold test we can confirm the authenticity of the $\sim\!4.617$ d
and $\sim\!1200$--1500\,d periodicities. The $\sim\!500$\,d feature, 
which in fact can be shown to be an alias (e.g. B99), fails the fold test.

These three features are the only significant periodicities in the frequency
spectrum of the original Lick data set. Additional significant features
are present in the periodogram of the new Lick data set. The most
prominent are located at $\sim\!141$\,d, $\sim\!14$\,d, and $\sim\!230$\,d.
They all fail the fold test.
The periodogram of the AFOE data set also shows additional significant peaks.
The most prominent peak is located at $\sim\!29$\,d
and is affirmed by the fold test. In addition,
peaks at $\sim\!145$\,d and $\sim\!245$\,d, the locations close to those
identified on the periodogram for the new Lick data set, are present,
but they fail the fold test.

This preliminary analysis of radial velocity signal from $\upsilon$ And
indicates existence of two periodicities, one at $\sim\!4.617$\,d,
and another at $\sim\!1200$ to $\sim\!1500$\,d.
This conclusion holds for all three data sets.
Therefore, we can confidently postulate that the radial velocity signal from
$\upsilon$ And is due to the motion of the star caused by the existence of
two companions, having orbital periods of $\sim\!4.617$\,d and
$\sim\!1200-1500$\,d.
A model consisting of three companions cannot be confidently postulated
on the basis of the frequency spectra of the radial velocity signal.
However, if such a model is postulated, the third companion should have
a period of either $\sim\!145$\,d or $\sim\!230$--245\,d in order to be
consistent with both the new Lick data and the AFOE data. 
In the AFOE data set, a periodicity of $\sim29$ d can be positively identified,
but it is not detected in the other data sets;
it would be interesting to understand its origin.
\section{Models}
Assume a model consisting of a single companion, labeled B, orbiting
$\upsilon$ And. Such a one-companion model predicts the radial velocity,
$V_{\rm mod,B}(t)$, at any given instant of time. 
The radial velocity signal due to the orbital motion of the star caused by
gravitational interaction with a companion is given by the following expression,
\begin{equation}
V_{\rm mod,B}=K \left[ \cos (f+\omega) + e \cos \omega \right ],
\label{eqn1}
\end{equation}
where $e$ is the eccentricity of the orbit, $f$ is true anomaly,
and $\omega$ is its argument of the periastron. The semi-amplitude
$K$ is proportional to the projected mass of the companion, $m \sin i$,
where $i$ is the angle between an observer's line-of-sight to a star
and the normal to the orbital plane of the companion.

The true anomaly, $f$, can be expressed in terms of the eccentric anomaly, $u$,
\begin{equation}
\tan \frac{f}{2} =\sqrt{\frac{1+e}{1-e}} \tan \frac{u}{2}.
\label{eqn2}
\end{equation}
In turn, eccentric anomaly, $u$, can be linked to time by means of Kepler's
equation,
\begin{equation}
\frac{2 \pi}{P} (t-T_{peri}) = u - e \sin u,
\label{eqn3}
\end{equation}
where $P$ is the period of the companion's orbit, and $T_{peri}$ is the time of 
periastron passage.

Equations (\ref{eqn1}) to (\ref{eqn3}) completely define the one-companion
model, giving the time dependence (albeit in an implicit form) of radial
velocity. There are five free parameters in this model: $K$, $P$, $e$,
$T_{peri}$, and $\omega$. If we assume two companions, labeled B and D,
to orbit $\upsilon$ And, then, in the first approximation, the two-companion
model is simply given by $V_{\rm mod}(t)=V_{\rm mod,B}(t) +V_{\rm mod,D}(t)$
with individual contributions given by (\ref{eqn1}).
There are 10 free parameters in the two-companion model.
The generalization to a model with an arbitrary number of companions is
straightforward. Thus, presupposing that the radial velocity signal from
$\upsilon$ And is mostly due to gravitational interactions with multiple
companions, the $N$-companion model can be written as follows
\begin{equation}
V_{\rm mod}(t)=\sum_{i=1}^{N} V_{{\rm mod},i}(t) + \mathcal{R}(t),
\label{eqn4}
\end{equation}
where $\mathcal{R}(t)$ encapsulates sources of radial velocity signal that
cannot be attributed to the presence of companions, but instead are intrinsic
to the star. They may, in principle, include pulsation and effects due to
the inhomogeneous and dynamic nature of the stellar convective and magnetic
patterns. However, in the case of $\upsilon$ And, there are arguments against
pulsations (B99), leaving convective inhomogeneities as the most likely source
of $\mathcal{R}(t)$.

The surface of a star having a convective zone is inhomogeneous in terms of
magnetic field, brightness, as well as vertical motion.  These inhomogeneities
occur on a variety of length scales and are transient.  This phenomenon alone
leads to variability of the radial velocity measured from the disk integrated
light. Such a variability is referred to as a ``jitter."
On short time scales the jitter is intrinsically stochastic.
Observations and theoretical arguments can be used to estimate the magnitude
of the short-term jitter, but not the actual form of $\mathcal{R}(t)$.
On long time scales the jitter should be modulated by the dependence of stellar
photospheric activity on possible cycles of the large-scale stellar magnetic
field.  Thus, the long-term character of $\mathcal{R}(t)$ should be sinusoidal.

For $\upsilon$ And, B99 quote the magnitude of the short-term jitter to be
$\sim\!10$\,m\,s$^{-1}$. This estimate is based on the work of
\citet{saa98} who investigated the relationship between the variability
of the radial velocity signal (i.e., jitter, unless the star has companions)
and various stellar properties for 72 stars in the Lick survey.
They established empirical relations, defined as the best power-law fits,
between the variability, $\sigma_{V}$, and quantities such as $B-V$ color,
stellar rotation period, $v \sin i$, and the fractional Ca II H \& K flux.
However, inspection of figures 1 and 2 of \citet{saa98} shows large
scatter of actual data around the empirical relations.
Thus, the value $10$\,m\,s$^{-1}$ is only a rough estimate of $\upsilon$ And's
jitter; values as large as $\sim\!20$\,m\,s$^{-1}$ cannot be ruled out on the
basis of Saar et al.'s diagrams.

Keplerian models of the radial velocity signal are defined by
$\mathcal{R}(t)=0$. Because the jitter is unavoidable, a keplerian model is
always incomplete and does not reflect accurately the reality. Thus, we
{\it should not} expect the keplerian model to fit the data accurately within
the known instrumental errors. The long-term modulation of the jitter,
if present, should be picked up by the keplerian model as a ``companion",
provided that the period of such modulations is short
enough and its amplitude is strong enough.

\section{Fitting methods and procedures}

We assume that the radial velocity signal from $\upsilon$ And is caused
by the presence of companions and thus adopt a keplerian model given by
(\ref{eqn4}) with $\mathcal{R}(t)=0$. We use $\chi^2$ as a merit function
to determine values of best-fit parameters:
\begin{equation}
\chi^2 =\sum_{k=1}^{M}
\left(\frac {V_k - V_{\rm mod}(t_k;a_1,\cdots,a_{5N})} {\sigma_k}\right)^2,
\label{eqn5}
\end{equation}
where $(t_k, V_k, \sigma_k)$, $k=1, \cdots, M$ are observational records for
a given data set with $M$ observations. Note that we use instrumental errors,
$\sigma_k$, as weights in the definition of $\chi^2$, leaving the jitter
unaccounted for. The function $\chi^2$ depends on $5N$ free parameters,
where $N$ is the number of companions in the model. We find the best-fit
parameters by minimizing the merit function $\chi^2$ with respect to all
$5N$ parameters. 

To minimize $\chi^2$ (Eqn.~\ref{eqn5}) we employ two different methods, the
Levenberg-Marquardt (LM) method, and the genetic algorithm (GA). The LM method
is an algorithm based on the concept of steepest gradient descent. 
The implementation of the LM algorithm can be found in \citet{pre92}.
The disadvantage of the LM method is that it needs a starting point
(set of parameters) and finds only a {\it local} minimum closest to the
starting point.  In the context of $\upsilon$ And, using the LM method
{\it requires} a prior calculation of signal's frequency spectrum to determine
starting values of orbital periods of companions. We have found the
LM method to be highly sensitive to the starting values of orbital periods,
but much less sensitive to the starting values of other parameters.
On occasion, the $\chi^2$ function has such a ``rugged landscape" 
that the LM method fails completely. We have found this to be the case when
trying to fit a two-companion model to the AFOE data set.

The GA method works by analogy with evolutionary updates of the genome.
It uses an ensemble (a population) of sets of parameters which evolves
exploring many options in parallel. The GA can find a global minimum
regardless of how rugged the landscape of the $\chi^2$ function is.
Because there is no single starting point in the GA, the prior calculation
of the signal's frequency spectrum is not necessary.
The disadvantage of the GA is its inefficiency in obtaining a very accurate 
(as opposed to approximate) best-fit solution.
Also, the GA provides no obvious protocol for calculating uncertainties
of the located best-fit solution.
Thus, the LM and the GA methods complement each other.
The GA algorithm can be designed to find not only the global minimum,
but also a number of other ranked minima. This is important because the
close second or third best-fit solutions may be physically as plausible
as the actual best-fit solution. Description of fundamentals and subtleties
of applying the GA  can be found in \citet{mic96}.
Applications of the GA in astronomy and astrophysics are discussed by
\citet{cha95}. For our calculations we used the GENOCOP system 
\citep{mic96}, which is the floating point implementation of the GA.
Our typical run had a population size of 70--100 and evolved for 5000--10000
generations. We ran 30--40 separate experiments on each model--data set
combination.

\section{Best fit Keplerian models}

\subsection{Two-companion models}

The frequency spectra of the radial velocity signal from $\upsilon$ And
(section 2) indicates two periodic components suggesting a two-companion
keplerian model. We label the two putative companions B and D,
for consistency with B99. In our model each companion is characterized by
5 parameters, thus the two-companion model has 10 free parameters to be
fixed by minimization of $\chi^2$ (Eqn.~\ref{eqn5}).

Table \ref{tab1} summarizes the best-fit two-companion solutions we have
found for all three data sets. A description of each best-fit solution
is divided into three sections. The first section gives the overall
properties of the fit, the other two sections list values of the best-fit
parameters for B and D, respectively.
For the sake of compactness, we don't list values of uncertainties of
estimated parameters. Uncertainties are generally about the same as those 
in B99 because we tune all solutions using the LM method. 

In the properties section we first list the method used to obtain a given
solution. The LM method uses a starting point with orbital periods as
indicated by the respective frequency spectra.
LM/GA stands for the solution found using the LM method and
confirmed using the GA method. In this context, ``confirmation" means that
the GA method yields the solution ``similar" to that obtained by the LM
method. Moreover, using the GA solution as a starting point in the LM
method recovers the original LM solution. Second, the value of $\chi^2$
is listed, together with the value of $\chi_{red}^2= \chi^2/L$, where $L$
is the number of degrees of freedom (the number of observations, $M$,
minus the number of parameters to be fitted);
in the case of two-companion models $L=M-10$.
Last, the standard deviation of residuals, labeled as ``RMS of residuals"
is given. The residuals are the values of $V_k-V_{\rm mod}(t_k)$,
$k=1, \cdots, M$.

Overall the best-fit solutions are quite similar for all three data sets.
The LM method failed to find the best-fit, two-companion solution for the
AFOE data.  The GA method yields several solutions of comparable ``fitness"
that can be grouped into two distinct categories.
For the AFOE data, Table 1 lists the fittest solution in each category.
There are some systematic differences between fits to Lick and AFOE data
sets, especially with regard to companion D.
Fig.~\ref{fig2} shows observed radial velocities together with their
two-companion best-fit models.
For compactness, this figure as well as Figs.~\ref{fig3}--\ref{fig4} cover
the period between 1992 and 2000, and do not show five earlier Lick data
points. However, {\em all} observations are used to obtain the best-fit
solutions.   It is quite clear from even a visual inspection of
Fig.~\ref{fig2} that the two-companion model does not fit the data well.     

It is expected that the value of  $\chi_{red}^2\approx 1$ for a good fit.
The values of $\chi_{red}^2$ in Table \ref{tab1} are in the range from
6.41 to 23.66. This seems to suggest that the two-companion model offers a bad
fit to the data. However, note that the $\chi_{red}^2\approx 1$ criterion for
the goodness of fit assumes completeness of the model.  Any keplerian model
is an incomplete model because the stellar jitter is not incorporated into it.
Thus, the best-fit solution should not be characterized by
$\chi_{red}^2\approx 1$, unless the $\sigma_{\rm jitter} \le \sigma_{\rm inst}$,
where $\sigma_{\rm jitter}$ and $\sigma_{\rm inst}$ are standard deviations of
the jitter signal and an average instrumental error, respectively.
The RMS of residuals is in the range from
28.3\,m\,s$^{-1}$ to 35.07\,m\,s$^{-1}$,
much higher than $\sim$10\,m\,s$^{-1}$ expected if the residuals were due to
instrumental errors alone. This indicates a bad fit unless the stellar jitter
is about 26 -- 34\,m\,s$^{-1}$. These are much higher values than 10\,m\,s$^{-1}$
adopted by B99, but cannot be definitively excluded on the basis of empirical
diagrams of \citet{saa98} as discussed in Sect.~3.

Fig.~\ref{fig1} (right column) shows the frequency spectrum of residuals left
after subtracting the best-fit, two-companion model from the signal.
Spectral features common to all three data sets exist and indicate the
existence of periodic components in the residuals. The prominent peaks are
at $\sim\!145$\,d and $\sim\!240$ d. Note that frequency spectrum of residuals
left after subtracting the formal best-fit model to the AFOE data shows no
features and is inconsistent with Lick data sets results.
On the other hand, the ``good" fit to the AFOE data leaves residuals with
frequency spectra consistent with those produced by the best fits to the
Lick data sets, except for an additional periodicity at $\sim\!29$\,d
present in the AFOE residuals.

{\it The apparent failure of two-companion models to fit well the data does
not, by itself, necessarily point out to the existence of the third companion;
instead, it may reflect a presence of the large but feasible jitter.
It is the existence of periodic component(s) in the residuals of the
two-companion model, rather than the large values of the residuals and
of $\chi_{red}^2$, that suggests an additional companion(s).}

\subsection{Three-companion models}

We now consider the keplerian model with three companions labeled
B, C, and D, from innermost to outermost. Such a model is characterized
by 15 parameters. Table \ref{tab2} summarizes the assorted three-companion
solutions for all three data sets. All fits were obtained by minimizing
$\chi^2$ (Eqn.~\ref{eqn5}) with respect to 14 parameters,
the period of the innermost companion, $P_{\rm B}$, having been {\em fixed}
for reasons of computational efficiency.
This is justified because the periodograms give the value of
$P_{\rm B}=4.6171$\,d with high accuracy. 

For each data set four different categories of solutions are listed.
The first is obtained by the LM method starting with $P_{\rm D}$ given
by the best-fit, two-companion solution, and $P_{\rm C}$ equal to 240 d
as indicated by the highest peak on the periodogram of residuals left
after subtracting the best-fit, two-companion model from the data.
Solutions in this category are the overall best-fits.
Hereafter we refer to them as the BF solutions.
The second category (hereafter referred to as the PC145 solutions) 
is obtained by the LM method starting with $P_{\rm D}$ given by the
best-fit, two-companion solution, and $P_{\rm C}$ equal to 145 days as
indicated by the second highest peak on the periodogram of residuals left
after subtracting the best-fit, two-companion model from the data.
The third category (labeled as the SE solutions) are the best-fit
solutions subject to the condition that eccentricities of all orbits are
$\le 0.1$, and the fourth category (labeled as the SEBC solutions) are
the best-fit solutions subject to the condition that eccentricities of
B and C orbits are $\le 0.1$.
The latter two models, SE and SEBC, were motivated by dynamical
stability considerations.
The description of each solution in Table \ref{tab2} is divided into four
sections, the first section gives the properties of the fit, the remaining
three sections list values of parameters for companions B, C and D.

Fig.~\ref{fig3} shows the original Lick data together with the four fits
listed in Table \ref{tab2}. Visual inspection of Fig.~\ref{fig3} suggests
that BF, SEBC, and SE solutions offer comparably good fits to the data,
whereas the PC145 solution provides a slightly worse fit. This impression is
confirmed by the values of $\chi^2$ in Table \ref{tab2}. The RMS of residuals
is in the range from $\sim\!16.6$ m\,s$^{-1}$ for the BF solution, to 
$\sim\!21$ m\,s$^{-1}$ for other solutions. Thus, the BF solution  offers a 
good fit providing that the jitter is gaussian with
$\sigma_{\rm jitter} \ge 13$\,m\,s$^{-1}$,
and other solutions offer a good fit providing that
$\sigma_{\rm jitter} \ge 18.5$\,m\,s$^{-1}$.
In this context, ``good fit" means that $\chi_{red}^2$, recalculated with
weights $\sigma_k =\sqrt{\sigma_k^2 +\sigma_{\rm jitter}^2}$,
has a value of approximately one.
As discussed in Sect.~3, it is plausible that $\upsilon$ And has random 
short-term jitter with magnitude equal to or even larger than required for
all four solutions to be ``good fits." 
However, in the presence of jitter characterized by
$\sigma_{\rm jitter} \ge 18.5$\,m\,s$^{-1}$
it would be unlikely to find a three-companion fit with the value of
$\chi^2$ as small as that we have found for the BF model. Therefore, it is
likely that the magnitude of $\upsilon$ And jitter is set by the fitness of
the BF model, and, consequently, the BF model is indeed the best solution
among the four considered here. We have calculated periodograms (not
shown) of residuals left after subtracting the three-companion models from
the signal. Periodograms of residuals for BF, SE, and SEBC models have no
features, indicating that these residuals are noise. The periodogram
of residuals for the PC145 model have peaks at $\sim\!79$ and $\sim\!243$\,days.
This may indicate that the PC145 model does not account for all periodic
components in the signal.

Fig.~\ref{fig4} shows the new Lick data together with BF, SE, SEBC, and
PC145 fits to that data set. The overall character of all solutions is
similar to that of analogous fits to the original Lick data. The 
relative fitness of different solutions, as measured by the value of
$\chi_{red}^2$, has changed. The fitness of the BF solution has increased,
whereas the fitness of all other solutions has decreased.
Periodograms of residuals for the BF and the SEBC models show no features,
whereas periodograms of residuals for the SE and, in particular,
the PC145 models show some statistically significant peaks.
 
Fig.~\ref{fig5} shows the AFOE data together with BF, SE, SEBC, and PC145 fits
to that data set. These solutions are generally fitter (have smaller values of
$\chi_{red}^2$) than their Lick counterparts.  This can be partially explained
by larger instrumental errors of the AFOE data.  However, the AFOE fits also
leave residuals with smaller values of RMS than their Lick counterparts.
This suggests that the AFOE data can be fitted slightly better by the
three-companion model than the Lick data.  The only model that leaves
residuals with possible periodic components is the SEBC.

The character of the solution in each category is consistent amongst all
three data sets. In addition, the only significant difference between the
BF, SEBC, and SE solutions are the eccentricities.
The fits to the AFOE data systematically yield a longer period and a larger
value of $K$ for companion D than the fits to the Lick data.
Our best fit to the original Lick data is very similar to that published
in B99, and our best fit to the the AFOE data is virtually identical
to that published in B99.
The new Lick data suggests that future data will not support the PC145 
and SE models; the BF remains the best model and the SEBC remains a
viable model.

{\it Three-companion models offer good fits to the data. The periods
and amplitudes of all companions, as well as eccentricities for
companions B and D, are well constrained by the existing data. However, the
eccentricity for companion C is not well constrained by the present data.}

\section{Dynamical stability}

In a Keplerian model, radial velocity data determine five parameters for
each companion, $(K,P,e,T_{peri},\omega)$.
The amplitude $K$ is related to the masses and orbital parameters as follows:
\begin{equation}
K = {m\sin i\over M_\star+m}\Bigg[{G(M_\star+m)\over a(1-e^2)}\Bigg]^{1/2},
\label{eqn6}
\end{equation}
where $M_\star$ and $m$ are the stellar and companion mass, respectively,
$G$ is the universal constant of gravitation, and $a$ is the orbital
semimajor axis (related to the orbital period $P$ through Kepler's third law).

From these parameters, we can calculate $m \sin i$ and $a$ for each companion,
provided the stellar mass $M_\star$ is known.  The mass of $\upsilon$ And
is estimated to be 1.2---1.4 $M_\odot$ \citep{for99};
following B99, we adopt $M_\star=1.3\,M_\odot$.
From the best-fit models for the new Lick observations and for the AFOE
observations (Table \ref{tab2}), the sets of parameters needed for orbital
dynamics studies are given in Table \ref{tab3}.
Note that for each of the companions, two orbital parameters ---
the inclination and the longitude of ascending node
--- remain undetermined by the radial velocity data.

Although the current estimated orbits of the companions are spatially
well separated, the (minimum) masses of the companions and the orbital
eccentricities of the two outer companions are sufficiently large that
significant perturbation of the orbits can be expected due to the mutual
gravitational forces amongst the companions.  
This is illustrated in Fig.~6 where we show the results of a numerical
integration of the equations of motion for this 4-body system
including all the (point-mass, Newtonian) gravitational forces amongst them,
for the best-fit models to the Lick and the AFOE data.
(We used a standard second order mixed variable symplectic integrator
\citep{wis91}, with a step size of 0.2 days;
the total energy error in this integration is quasiperiodic and 
bounded to a few parts in $10^8$.)
In this integration, we assumed that the orbits are coplanar and edge-on
to the line-of-sight.  This assumption is not necessarily realistic
but it provides a useful fiducial case for measuring the effect of
departures from coplanarity and edge-on orientation (which we explore
further below).
The figure shows that the orbital semimajor axes are little perturbed
(not unexpected, as all the orbital periods are well separated).
However, a remarkable feature of the evolution is that the orbital
eccentricities of all companions are perturbed significantly on relatively
short timescales.

The middle companion, C, exhibits the most dramatic perturbation,
its eccentricity varying periodically from a maximum ($\sim\!0.35$
in the Lick best-fit model, $\sim\!0.28$ in the AFOE best-fit model)
to a minimum near zero; companion D's eccentricity exhibits a variation
with the same period but much smaller amplitude.
The period of these variations is about 7000 yr for the Lick model,
and about 3500 yr for the AFOE model.
These eccentricity variations (and corresponding apsidal variations)
arise due to a secular interaction between the outer two companions.
(See, for example, \citet{bro61}.)
This interaction can be described approximately as a superposition of
two eigenmodes for the evolution of the ``eccentricity vector'',
$(e\cos\omega,e\sin\omega)$, for each of the companions C and D.
The outer companion D's apsidal rate is dominated by the slowest frequency mode.
For the middle companion C, the two modes have nearly equal amplitudes
(so that the magnitude of the eccentricity nearly vanishes periodically),
and its apsidal motion is limited to the range --90 deg to +90 deg
relative to the apsidal line of companion D.
In this context, it is noteworthy that the radial velocity data do not
constrain very well the eccentricity of companion C;
good fits to the data include models with small values of $e_C$
(cf.~discussion in the previous section).
Interestingly, we have found that the large amplitude oscillation of the
eccentricity of C persists in the ``good fit'' (SEBC) model as well.

In Fig.~6,  we see that in the Lick model the innermost companion, B,
also suffers a dramatic eccentricity variation, albeit on a longer timescale;
however, we consider that this is not ``real'' because the proximity of
companion B to the star would subject it to general relativistic precession
that would dominate its secular evolution, suppressing the amplitude of the
eccentricity perturbations (see \citet{riviera00}).

The outer two companions are the most strongly coupled
and companion B provides only a very small perturbation to their orbital
evolution.
It is important to note that the mutual gravitational interactions
of the outer two companions is sensitive to their unknown orbital
inclinations and relative orientation of their lines of nodes,
{\it i.e.}, $i_C, i_D$ and $\Omega_D-\Omega_C$.
For given values of these parameters, the relative inclination of the two
orbits, $\phi_{{}_{CD}}$, is given by
\begin{equation}
\cos\phi_{{}_{CD}} = \cos i_C \cos i_D
                    + \sin i_C \sin i_D \cos(\Omega_D-\Omega_C).
\label{eqn7}
\end{equation}
The uncertainties in the other known parameters will also affect the dynamics
and stability of the system.  Thus, in principle, there is a very large
volume of parameter space that needs to be investigated for dynamical studies.
Here we confine our discussion to a subset of this parameter space related
to the undetermined parameters only. 

As the long term dynamics and stability of the system is determined largely
by the mutual gravitational interactions of the outer two companions,
in the numerical investigations described below, we have neglected the presence
of the innermost companion. This allows us to use larger integration step sizes
(we used an 8 day step size) so that a large number of integrations can be
completed with relatively modest computer resources.
These calculations provide a necessary but not sufficient condition for
the stability of the full system.
 
We have investigated the orbital evolution of the 3-body system,
$\upsilon$ And with companions C and D, for a range of these unknown parameters
by numerically integrating the orbits for timescales ranging from $10^6$ yr
to 1 byr.  ($\upsilon$ And's age is estimated to be 1.6---4.7 byr
\citep{for99}.)
We find that for values of companion masses not too much larger than the
minimum and for small values of relative inclinations, the orbital evolution is
quite regular and the orbital parameters exhibit only periodic variations of
constant amplitude:
the qualitative character of the eccentricity evolution is similar to that 
described above for the strictly coplanar, edge-on case (cf.,~Fig.~6);
the orbital inclinations also vary periodically with significant amplitude
about the initial values, but with a relatively small amplitude periodic
variation of the {\it relative} inclination.
However, for large-relative-inclinations the evolution is chaotic:
the amplitude and frequency of the orbital variations is erratic.
Fig.~7 shows two representative examples for the case where the masses of
C and D are each twice their minimum mass (i.e., their orbits are inclined
30 deg to the plane of the sky); the relative orbital inclination is small,
7.5 deg, in one case, and moderately large, 41.4 deg, in the other.

Several measures can be used to quantify the orbital stability of this system.
Commonly used measures include the maximal Lyapunov exponent (which measures
the rate of exponential divergence of nearby trajectories in phase space),
and time-evolution of the fundamental frequencies of the system.
Here we have chosen to employ the {\it range} of variation of companion C's
eccentricity as a stability measure; this provides a very direct visualization
of the orbital stability as a function of the orbital inclinations, as 
shown below.

Figure 8 summarizes the results of a large suite of such numerical 
integrations, for three choices of initial values of $i_C$ and $i_D$
that span the range of companion masses up to about 8-15 Jupiter masses. 
Each panel in this figure plots the range of variation of companion C's
orbital eccentricity, $e_C$, as a function of the relative orientation
of the line of nodes, $(\Omega_D-\Omega_C)$
(equivalently, the relative orbital inclination of C and D,
$\phi_{{}_{CD}}$) over a timespan of $10^6$ yr.
This timespan  is long enough that the strongest instabilities
would be easily detected, and short enough that we can cover a
wide range of the unknown orbital parameters.

Qualitatively, we find that the range of $e_C$ is modest and stable for
small values of $(\Omega_D-\Omega_C)$ and $\phi_{{}_{CD}}$ but increases
dramatically for larger values of these parameters.
Indeed, the quasiperiodic character of the orbital evolution is possible
only for a limited range of $\phi_{{}_{CD}}$, which decreases 
for decreasing values of $\sin i$.
(Outside these ranges the orbits are chaotic, and close encounters between
C and D could occur on timescales much less than the age of the system.)
We find that the parameter range for stable orbits is significantly smaller
for the AFOE best-fit model compared to the Lick best-fit model.
For the case of strictly coplanar orbits, the Lick best-fit model
is unstable for $\sin i \ltsim 0.23$ whereas the AFOE best-fit model is
unstable for $\sin i \ltsim 0.33$.
For the Lick data, we find quasiperiodic orbital evolution for
$\phi_{{}_{CD}}\ltsim 55^\circ$ $\ltsim35^\circ$, and $\ltsim10^\circ$
for initial $\sin i_C=\sin i_D = 0.9$, $0.5$ and $0.25$, respectively.
For the AFOE data, we find quasiperiodic orbital evolution for
$\phi_{{}_{CD}}\ltsim 30^\circ$ and $\ltsim15^\circ$
for initial $\sin i_C=\sin i_D = 0.9$ and $0.5$ respectively;
no orbital stability is found possible for $\sin i\leq1/3$.
We note that these estimates are based on our most comprehensive set of
$10^6$ yr numerical integrations. 
A suite of $10^8$ yr integrations near the stability boundaries estimated
above finds signs of weak chaos in a few cases.
We expect that on longer timescales, the dynamically stable range
of parameters would be smaller; the preceding estimates provide
an upper limit for the dynamically stable range of the unknown
parameters (inclinations and nodes).

To summarize:
(i) the range of parameter space where long term orbital stability is
possible is quite sensitive to the orbital parameters;
and (ii) the requirement of long term orbital stability constrains the
orbital inclinations of C and D to values of $\sin i_C, \sin i_D >0.23$;
this requirement also provides meaningful constraints on the relative
orbital inclinations and longitudes of nodes.

\section{Discussion and Conclusions} 

\subsection{Models}

We have analyzed the radial velocity data given in B99 on $\upsilon$ And,
as well as more recent unpublished data on $\upsilon$ And taken by the
Lick Observatory group. Our motivation for re-examining the B99 data was
the prospect of finding feasible (in the $\chi^2$ sense) models that would
be more dynamically stable than those published in B99.
Our methodology differed from that of B99 in the following aspects. 

First, we explicitly did not analyze
what is referred to in B99 as the combined data set.
We do so for two reasons.  One is that combining data sets taken with different 
instruments incurs some risk of introducing systematics into the combined data.
More importantly, the existence of two independent data sets serves the
valuable purpose of corroborating one set of data by the other.
This corroborative aspect is important to maintain when there are few systems
capable of measurements with high precision. 
In the longer term, a continuing set of independent observations could reveal
the presence of systematics in one of the systems should they develop.
Finally, to the extent that any model of the data describes physical reality,
the model should fit all data sets.  Inconsistencies between data sets in this
regard should be treated as a warning that the model may be incomplete or
incorrect.  We would encourage observers to maintain this important
independence in their data sets, while recognizing the inherent strengths
(such as better characterization of candidate models) of a larger
combined data set.

Second, in finding the minima of the $\chi^2$ function we relied on a
{\it global} search method -- the genetic algorithm. In cases where there is
reason to suspect that there may be multiple periods in the data,
the $\chi^2$ function is likely to be complicated. Such functions are referred
to as having a ``rugged landscape" because they have a large number of
intermingled local minima and maxima.  The genetic algorithm will find the
global minimum (and other ranking minima) even for functions with an extremely
rugged landscape. This is in contrast to the standard, steepest gradient
descent, minimization method which is {\em local} and converges to a local
minimum located closest to the starting values.  Thus, the local method does
not guarantee finding the global minimum.  Moreover, as starting values of
periods can only be provided from data frequency spectrum, the local method
requires estimating and interpreting such a spectrum.  Extracting periodicities
from the data spectrum may be a laborious process not free of ambiguity,
as documented by efforts in B99 and also in our present paper.
For example, consider differences in frequency spectra between different
data sets (Fig.~\ref{fig1}). A good illustration of the difficulties in
extracting periodicities from the frequency spectrum is provided by the
inconsistency of the spectra of residuals of two-companion fits to the AFOE
data.
(We note that B99 reported checking the uniqueness of their
three-companion model with the genetic algorithm, although no details
were given.)
These problems can be entirely avoided by using the global minimization
technique. In the present paper, in addition to using the genetic algorithm,
we also employed the local method, both for continuity with B99 and also as
a means of fine-tuning the global solution.
We recommend that future
studies avoid relying exclusively on the local minimization technique.

Third, instead of looking for the ``best-fit" model we looked for a {\it set}
of viable models. We did not specify formal criteria for model ``viability",
but very roughly, the ``good-fit" models
are characterized by:
(a) fitness that is not much smaller than the fitness of the BF model, and
(b) residuals with no periodic components.

We have found that two-companion models are not viable, primarily because they
leave periodic residuals. We have identified four potentially viable
three-companion models.  Our best-fit, three-companion Keplerian model is
consistent with that of B99.  There are differences in details of the orbital
parameters, but nothing fundamental.
The only significant difference between the best-fit model and the
second-best-fit model (SEBC) is the small eccentricity of companion C's orbit.
Interestingly, the long term dynamical behavior of the SEBC model is 
qualitatively similar to the best-fit model. 
The third-best-fit model (SE) has all three companions on orbits with small
eccentricities. Finally, we have found that a model (PC145) with companion C
orbiting the star with a period of 145 days also offers a good fit to the data.
These last two models describe a system which is potentially more stable than 
that described by the BF model, but the BF model is most consistent with the data.

The new Lick data provides further strong support for the BF model.
This is the only model for which the value of $\chi^2_{red}$ has decreased with
addition of the new data, as expected for a physically viable model.
Although the value of $\chi^2_{red}$ for the SEBC model has increased slightly,
this model still deserves being monitored against future data.	
Values of $\chi^2_{red}$ for SE and PC145 models increased significantly,
practically eliminating these models from consideration.
We conclude that parameters of the three-companion model of the $\upsilon$ And
radial velocity data are well constrained by the data, with the exception of
the eccentricity of the middle companion. Regardless of the eccentricity of the
middle companion the system described by such a model presents a challenge
from the point of view of dynamical stability. 

Because of dynamical stability concerns and also because the present data
allows the $\sim\!240$\,d periodicity to be almost sinusoidal, one can
contemplate a model consisting of two companions, B and D, and a stellar
cycle with a period of $\sim\!240$\,d causing a radial velocity variation of
about 50\,m\,s$^{-1}$.  There is some evidence that magnetic cycles can indeed
induce periodic changes in integrated light velocity.
\citet{dem94} reported a sinusoidal variation of solar integrated
light velocity with semiamplitude of 14\,m\,s$^{-1}$ and period of 11 years.
This variation is positively correlated with the solar disk-averaged
magnetograph signal and is over three times as large as the Sun's jitter of
4\,m\,s$^{-1}$. However, \citet{mcm93}, using a different technique,
have not found any long-term variability in solar integrated light velocity.
Thus, at present such a model is speculation, especially as it is not clear
that a stellar magnetic cycle with a sub-yearly period is possible
and capable of causing light velocity variations as large as 50\,m\,s$^{-1}$
which would be necessary to explain the $\upsilon$ And observations.

\subsection{Orbital dynamics constraints}

The radial velocity data do not provide constraints on all the orbital
elements of the putative companions: the orbital inclinations to the
plane of the sky and the orientation of the lines of nodes in the
plane of the sky remain undetermined.

Direct constraints on the viewing geometry of the $\upsilon$ And system are few.
The fact that there is no transit phenomenon \citep{henry00},
as has recently been found for the star HD209458 \citep{cha00} ,
implies that the orbital inclination of the inner companion must be
$\ltsim 82.8$ degrees.
Additional constraints on the inclination of the outer companion's orbit
come from observations of $\upsilon$ And by Hipparcos. Using $2\sigma$
detection limits for Hipparcos accuracy for relative astrometric observations,
which is roughly 2 milliarcseconds \citep{perryman96},
the absence of a reliable detection of a perturbation due to a companion with
the orbital periods given by the best-fits gives an upper limit to the true
companion masses of 27.7 $M_{\rm J}$ and 25.12 $M_{\rm J}$ for the new Lick
and AFOE data respectively.
Using the minimum masses for the corresponding best-fits,
$\sim 3.76$ $M_{\rm J}$ and $\sim 4.92$ $M_{\rm J}$,
the absence of astrometric detection means that the inclination must be
$\gtsim 7.8$ degrees ($\sin i \gtsim 0.136$) and $\gtsim 11.3$ degree
($\sin i \gtsim 0.196$) for the new Lick and AFOE data respectively.
(Note that the minimum masses given for companion D in B99, Table 3,
columns 2 and 3, are in error.  They are too high by a factor of
${(1-e^2)}^{-1/2}$.)

\citet{maz99} have used a combination of Hipparcos data and the
spectroscopically-determined best-fit orbital elements from B99 to deduce a
semi-major axis for the orbit of the outer companion and hence an estimate
of the true mass.  (This is not inconsistent with the fact that Hipparcos did
not detect an astrometric perturbation for $\upsilon$ And.)
The $2\sigma$ mass range from Mazeh et al.~is $10.1^{+9.5}_{-6.0}$ $M_{\rm J}$.
This approach is useful, but it depends upon the assumed set of spectroscopic
elements. The spectroscopic elements we find here for the best-fit model differ
somewhat from those used by Mazeh et al.  The robustness of the derived mass
against variations in these elements is unknown.  Taken at face value,
this mass would constrain the orbital inclination of the outer companion,
at the $2\sigma$ level, to be $\gtsim 11$ degrees (14.5 degrees) for the
new Lick (AFOE) data.

Here we have used orbital dynamics and stability considerations to infer
constraints on the undetermined orbital parameters, using the fact that the
mutual gravitational interactions in the system depend both upon the true
companion masses and on all the orbital elements.
{\it Our studies show that dynamical stability requires 
$\sin i \gtsim 0.23$ for the best-fit model to the New Lick data, 
and $\sin i \gtsim 0.33$ for the best-fit model to the AFOE data.}
The $i$ here refers to the present orbital inclinations of companions C and D,
assumed equal but not necessarily coplanar.
Thus the most massive companion, D, can be at most about 20 Jupiter masses.
Our studies also show that the range of allowed relative inclinations
of the orbits of the outer two companions C and D decreases with decreasing
$\sin i$ (that is, it is sensitive to the masses of the companions).
Even for values of $\sin i$ close to 1, their relative inclination can be no
greater than about 60 degrees for the Lick best-fit model
($\sim\!30$ degrees for the AFOE best-fit model).
Finally, our analysis provides constraints on the lines-of-nodes of the
orbits of companions C and D (Fig.~8).

The above conclusions are qualitatively consistent with those of \citet{lau99}
and \citet{riviera00} whose studies consisted of a survey of the long term
dynamical stability of B99's candidate models for the $\upsilon$ And system.
Laughlin and Adams used statistical arguments based on the long term behavior
of the orbital elements (of initially nearly coplanar orbits)
to infer likely values of the relative orbital inclination and orbital
eccentricities of the outer companions. (Note that these authors used an
erroneous definition of relative inclination, namely ``$i_C-i_D$''; the correct
definition is given in Eqn.~7.  It is unclear how their conclusions regarding
the likely relative inclinations would be affected if this error were
corrected.)  
Our focus here was on using stability considerations to quantitatively constrain
possible orbital geometries.  All three dynamical studies suggest that small
differences
in the parameters of the three-companion models lead to significant differences
in the dynamical stability of the system. This extreme sensitivity, while 
advantageous in constraining the nature of the system if one has a firm handle
on the parameters, raises questions as to how general the conclusions can
be when the parameters are not well known --- as is the case here.

Our analysis of the orbital evolution shows that for small relative inclinations
of the outer two companions C and D, the eccentricity of companion C varies
periodically from a maximum of $\sim\!0.3$ to a minimum of essentially zero
with a period of a few thousand years.
This peculiar behavior may harbor clues to the origin of the unusual orbital
arrangement in this system.
It is noteworthy that our analysis of the radial velocity data led to
the conclusion that the eccentricity of companion C is not very well
constrained by the observations: models with small eccentricities for
companion C provide nearly as good fits to the data (in a $\chi^2$ sense)
as the best-fit solutions.

\acknowledgments

We thank Goeff Marcy for providing the new Lick data for $\upsilon$ Andromedae,
and Tim Brown for suggesting improvements to our paper.
This research was conducted at the Lunar and Planetary Institute,
which is operated by the Universities Space Research Association under
contract No.~NASW-4574 with the National Aeronautics and Space Administration.
This is Lunar and Planetary Institute Contribution No.~1002.


\clearpage



\figcaption[fig1.eps]{Frequency spectra of the radial velocity data.
Different rows correspond to different data sets as indicated by labels.
The left column shows frequency spectra of actual radial velocities.
The right column shows frequency spectra of residuals of two-companion
best-fits. For the AFOE data frequency spectra of residuals for both,
the best and the good fits are shown.
Arrows points to locations of prominent peaks. \label{fig1}}

\figcaption[fig2.eps]{Best-fit, two-companion models to the radial velocity
signal from $\upsilon$ And as listed in Table \ref{tab1}.
Solid curves indicate modeled signal from the outer companion.
Data minus modeled signal from the inner companion are indicated by dots.
Original instrumental errors are indicated by vertical bars.
Inserts show the data minus the modeled signal from one companion folded with
the period of the remaining companion as indicated by the label.
On each of the inserts, the horizontal scale is one period, while the
vertical scale accommodates the amplitude, $K$, of that periodic component;
values of $K$ are listed in Table 1.
\label{fig2}}

\figcaption[fig3.eps]{Three-companion models, listed in Table \ref{tab2},
yielding good fits to the original Lick data set.
Solid curves indicate modeled signal from two outer companions.
Data minus modeled signal from the inner companion are indicated by dots.
Original instrumental errors are indicated by vertical bars.
Inserts show the data minus the signal from the model's two companions,
folded with the period of the remaining companion as indicated by the label.
On each of the inserts, the horizontal scale is one period, while the
vertical scale accommodates the amplitude, $K$, of that periodic component;
values of $K$ are listed in Table 2.
\label{fig3}}

\figcaption[fig4.eps]{Three-companion models, listed in Table \ref{tab2},
yielding good fits to the new Lick data set.
See also legend for Fig.~\ref{fig3}.
\label{fig4}}

\figcaption[fig5.eps]{Three-companion models, listed in Table \ref{tab2},
yielding good fits to the AFOE data set. See also legend for Fig.~\ref{fig3}.
\label{fig5}}

\clearpage

\figcaption[fig6.eps]{The orbital evolution of the three companions,
B (green), C (magenta) and D (blue),
for the best-fit model to the new Lick data (left)
and to the AFOE data (right),
assuming coplanar, edge-on orbits of all three companions.
The top panel shows the orbital semimajor axes and periastron and
apoastron distances; the middle panel shows the orbital eccentricities;
the bottom panel shows the arguments of periastron
(for companions B and C we show this angle relative to $\omega_D$).
\label{fig6}}

\figcaption[fig7.eps]{The orbital evolution of the two outer companions,
C and D, for the best-fit model to the new Lick data (Table 3), 
assuming initial $\sin(i_C)=\sin(i_D)=0.5$, and
(i) initial $\Omega_D-\Omega_C = 15^\circ$, initial relative orbital
inclination $7.5^\circ$ (on the left),
(ii) initial $\Omega_D-\Omega_C = 90^\circ$, initial relative orbital
inclination $41.4^\circ$ (on the right).
The top panel shows the semimajor axis, periastron distance and apoastron
distance; the middle panel shows the eccentricity; the bottom panel shows
the inclination to the plane of the sky. Elements for companion C are shown
in magenta and for companion D in blue; their relative orbital inclination
is shown as the black curve in the bottom panel.
\label{fig7}}

\figcaption[fig8.eps]{The range of eccentricity variation for companion C
over $10^6$ yr as a function of initial $(\Omega_C-\Omega_D$),
for the best-fit model to the new Lick data (left panels) and to the
AFOE data (right panels), for various values of initial
$\sin(i_C)=\sin(i_D)\equiv \sin i$.
Note that the horizontal scale is the same for initial $(\Omega_C-\Omega_D$)
in all panels, but not so for the relative inclination (indicated at the
top of each panel.
\label{fig8}}

\clearpage

\begin{table}
\centering
\caption{Best-fit two-companion models.
Values of $K$ are given in m\,s$^{-1}$, $P$ in days,
$T_{peri}$ in JD--2440000\,JD, and $\omega$ in degrees.}
\begin{tabular}{c c | c | c | c c}  \hline \hline
\multicolumn{2}{c|}{data sets}&{\bf Original Lick data}
&{\bf New Lick Data}&\multicolumn{2}{c}{\bf AFOE data} \\ \hline
\multicolumn{2}{c|}{solutions}&best fit&best fit&best fit&good fit\\ \hline
&&&&&\\
&method&LM/GA&LM/GA&GA&GA\\
&$\chi^2$&884.6&1868.8&269.2&296.9\\
&$\chi^2_{red}$&11.2&23.66&6.41&7.07\\
&RMS&30.75&35.07&28.3&30.52\\ 
&&&&&\\
&$K$&72.83&68.11&73.49&67.5\\
&$P$&4.6171&4.6171&4.6168&4.6174\\
{\bf B}&$e$&0.137&0.062&0.11&0.0\\
&$T_{peri}$&7494.26&7490.07&2551.28&11833.38\\
&$\omega$&8.92&45.09&133.18&291.97\\ 
&&&&&\\
&$K$&66.8&66.98&115.31&87.61\\
&$P$&1239.3&1216.67&1230.92&1598.59\\
{\bf D}&$e$&0.176&0.136&0.38&0.175\\
&$T_{peri}$&7218.62&7652.8&7173.43&4106.42\\
&$\omega$&137.51&260.53&123.5&355.06\\ \hline
\end{tabular}
\label{tab1}
\end{table}

\clearpage

\begin{table}
{\tiny
\caption{Best-fit three-companion models.
Values of $K$ are given in m\,s$^{-1}$, $P$ in days,
$T_{peri}$ in JD--2440000\,JD, and $\omega$ in degrees.
For clarity, we do not list the uncertainties in model parameters;
these are generally similar to those in B99, and may be obtained
from the authors.}
\begin{tabular}{c c| c c c c| c c c c| c c c c}  \hline \hline
\multicolumn{2}{c|}{data sets}&\multicolumn{4}{c|}{\bf Original Lick data}
&\multicolumn{4}{c|}{\bf New Lick Data}&\multicolumn{4}{c}{\bf AFOE data}
\\ \hline
\multicolumn{2}{c|}{solutions}&BF&SE&SEBC&PC145&BF&SE&SEBC&PC145
&BF&SE&SEBC&PC145\\ \hline
&&&&&&&&&&&&&\\
&method&LM/GA&GA&GA&LM/GA&LM/GA&GA&GA&LM/GA&LM/GA&GA&GA&GA\\
&$\chi^2$&242.1&323.1&302.0&433.5&256.6&604.1&465.1&904.8&42.5&106.3&54.8&85.1\\
&$\chi^2_{red}$&3.23&4.31&4.02&5.78&2.47&5.81&4.47&8.7&1.12&2.8&1.44&2.24\\
&RMS&16.58&21.22&20.94&21.03&14.56&21.5&18.09&26.35&12.19&18.14&14.43&15.79\\ 
&&&&&&&&&&&&&\\
&$K$&71.71&70.71&71.13&74.63&71.14&69.67&70.23&70.0&75.57&73.56&75.11&76.25\\
&$P$&4.6171&4.6171&4.6171&4.6171&4.6171&4.6171&4.6171&4.6171&4.6171&4.6171
&4.6171&4.6171\\
{\bf B}&$e$&0.075&0.069&0.074&0.1478&0.043&0.0&0.035&0.088&0.042&0.097&0.024
&0.017\\
&$T_{peri}$&10315.46&7624.46&1414.22&7494.42&10315.16&3134.12&4211.71&10315.64
&10315.97&1834.38&1862.34&5772.05\\
&$\omega$&25.11&85.75&66.89&24.11&0.58&247.88&27.44&36.58&64.95&66.89&87.29&13.14\\ 
&&&&&&&&&&&&&\\
&$K$&54.48&48.86&53.27&32.83&55.63&49.88&55.34&35.29&53.55&38.65&46.16&40.52\\
&$P$&241.17&241.47&242.08&145.67&241.46&240.12&240.85&142.81&243.38&239.23
&243.6&146.0\\
{\bf C}&$e$&0.331&0.1&0.1&0.204&0.35&0.1&0.05&0.20&0.23&0.1&0.0&0.5\\
&$T_{peri}$&11124.61&0.0&943.28&6951.61&11124.67&3923.28&8462.15&11234.3
&11114.39&8005.29&9254.59&11806.98\\
&$\omega$&250.54&218.58&223.55&28.36&248.21&254.68&232.56&238.39&219.21
&236.6&354.26&117.45\\ 
&&&&&&&&&&&&&\\
&$K$&64.63&65.73&67.81&66.14&64.47&61.57&68.26&60.9&85.09&74.43&91.9&99.25\\
&$P$&1276.19&1455.37&1439.6&1262.2&1291.79&1272.51&1309.5&1248.23&1490.09
&1441.62&1529.65&1210.51\\
{\bf D}&$e$&0.295&0.065&0.177&0.422&0.29&0.1&0.397&0.272&0.426&0.1&0.47&0.34\\
&$T_{peri}$&13853.84&4017.92&2781.22&7255.8&13894.1&6227.23&3457.34&13609.89
&14516.9&5.06&6993.81&10857.2\\
&$\omega$&246.8&206.78&243.35&167.84&242.97&253.36&240.33&199.22&245.63
&252.57&254.61&121.39\\ \hline
\end{tabular}
\label{tab2}
}
\end{table}

\clearpage

\begin{table}
\centering
\caption[]{Parameters and initial conditions for orbital dynamics studies.
For clarity, we do not list the uncertainties in model parameters;
these are generally similar to those in B99, and may be obtained
from the authors.}
\begin{tabular}{c c c c c c c}  \hline \hline
& Companions & $m \sin i$ & semimajor axis & eccentricity & $\omega$ 
& mean anomaly \\
&   & $10^{-3} M_{\odot}$ & AU &  & radians & radians \\ \hline
Lick & B & 0.657 & 0.0592 & 0.0430 & 0.010 & 4.651 \\
     & C & 1.804 & 0.8282 & 0.3478 & 4.332 & 2.149 \\
     & D & 3.732 & 2.5334 & 0.2906 & 4.241 & 6.192 \\ 
\noalign{\smallskip}
AFOE & B & 0.703 & 0.0592 & 0.042 & 1.114 & 3.552 \\
     & C & 1.819 & 0.8326 & 0.230  & 3.826 & 2.646 \\
     & D & 4.917 & 2.7865 & 0.426  & 4.287 & 6.087 \\ \hline
\end{tabular}
\label{tab3}
\tablenotetext{}{Orbital elements refer to Epoch $T_0$ = JD 2450000.0}

\end{table}


\begin{thebibliography}{}
\bibitem[Black and Scargle(1982)]{bla82}
Black, D.C., \& Scargle, J.D. 1982, ApJ, 263, 854

\bibitem[Butler et al.(1997)]{but97}
Butler, R.P., Marcy, G.W., Williams, E., Hauser, H., Shirts, P. 1997, ApJ 474, L115

\bibitem[Brouwer and Clemence(1961)]{bro61}
Brouwer, D., \& Clemence, G.M. 1961,
Methods of Celestial Mechanics (NewYork: Academic Press)

\bibitem[Butler et al.(1999)]{but99}
Butler, R.P., Marcy, G.W., Fisher, D.A., Brown, T.M, Contos, A.R., Korzennik, S.G.,
Nisenson, P., Noyes, R.W. 1999, ApJ, 526, 916

\bibitem[Charbonneau(1995)]{cha95}
Charbonneau, P., 1995, ApJS, 101, 309

\bibitem[Charbonneau(1999)]{cha00}
Charbonneau, D., Brown, T.M., Latham, D.W., Mayor, M. 2000, ApJ, 532, L55

\bibitem[Cumming(1999)]{cum99}
Cumming, A., Marcy, G.W., Butler, R.P. 1999, ApJ, 526, 890

\bibitem[Deming and Plymate(1994)]{dem94}
Deming, D., \& Plymate, C. 1994, ApJ, 426, 382


\bibitem[Ford(1999)]{for99}
Ford, E.B., Rasio, F.A., and Sills, A. 1999, ApJ, 514, 411

\bibitem[Henry et al. (2000)]{henry00}
Henry, G.H., Baliunas, S.L., Donahue, R.A., Fekel, F.C., Soon, W. 2000,
ApJ, 531, 415

\bibitem[Laughlin and Adams(1999)]{lau99}
Laughlin, G., \& Adams, F.C. 1999, ApJ, 526, 881

\bibitem[Lomb(1976)]{lom76}
Lomb, N.R. 1976, Ap\&SS, 39, 447

\bibitem[Marcy et al.(1999)]{mar99}
Marcy, G.W., Cochran, W.D., \& Mayor, M. 1999, To appear in Protostars and 
Planets IV, ed. V. Mannings, A. Boss, \& S. Russell, Univ.~of Arizona Press,
Tucson

\bibitem[Marcy and Butler(1998)]{mar98}
Marcy, G.W., \& Butler, R.P. 1998, ARAA, 36, 57

\bibitem[McMillan et al.(1993)]{mcm93}
McMillan, R.S., Moore, T.L., Perry, M.L., Smith, P.H. 1993, ApJ, 403, 801

\bibitem[Mayor et al.(1997)]{may97}
Mayor, M., Queloz, D., Udry, S., \& Halbwachs, J.-L. 1997, in IAU Colloq. 161, 
Bioastronomy 96, ed. C. B. Cosmovici, S. Bowyer, \& D. Werthimer (Bologna: 
Editride Compositori), 313

\bibitem[Mazeh et al.(1999)]{maz99}
Mazeh, T., Zucker, S., dalla Torre, A., \& van Leeuwen, F. 1999, ApJ, 522, L149

\bibitem[Michalewicz(1996)]{mic96}
Michalewicz, Z. 1996, Genetic Algorithms + Data Structures = Evolution Programs
(3th ed.; Berlin: Springer-Verlag)

\bibitem[Perryman et al. (1996)]{perryman96}
Perryman, M.A.C., Lindegren, L., Arenou, F., Bastian, U., Bernstein, H.H.,
van Leeuwen, F., Schrijver, H., Bernacca, P.L., Evans, D.W., Falin, J.L.,
Foreschle, M., Grenon, M., Hering, R., Høg, E., Kovalesvsky, J.,
Mignard, F., Murray, C.A., Penston, M.J., Peretsen, C.S., Le Poole, R.S.,
Söderhjelm, S., \& Turon, C. 1996, A\&A, 310, L21


\bibitem[Press et al.(1992)]{pre92}
Press, W.H., Teukolsky, S.A., Vetterling, W.T., \& Flannery, B.P. 1992,
Numerical Recipes:The Art of Scientific Computing
(2nd ed.; New York: Cambridge Univ. Press)

\bibitem[Riviera and Lissauer, 2000]{riviera00}
Riviera E.J., Lissauer, J.J. 2000, ApJ, 530, 454

\bibitem[Saar et al.(1998)]{saa98}
Saar, S.H., Butler, R.P, \& Marcy, G.W. 1998, ApJ, 403, L153

\bibitem[Scargle(1982)]{sca82}
Scargle, J.D. 1982, ApJ, 263, 835 

\bibitem[Vogt et al.(2000)]{vog00}
Vogt, S.S., Marcy, G.W., Butler, R.P, Apps, K. 2000, ApJ, 536, 902

\bibitem[Wisdom and Holman(1991)]{wis91}
Wisdom, J., \& Holman, M. 1991, AJ, 102, 1528
\end{thebibliography}
\end{document}